\newif\ifarxiv
\arxivtrue 

\ifarxiv
	\documentclass[12pt]{article}
	\usepackage[margin=1in]{geometry}
	\usepackage[slantedGreek]{mathpazo}
	\usepackage[pdftex]{graphicx}
	\usepackage{amsfonts}
	\usepackage{amssymb}
	\usepackage{amsthm}
	\usepackage{graphicx,float}
	\usepackage{amsmath}%
	\usepackage{hyperref}
	\usepackage{natbib}
	\usepackage{algorithm, algorithmic}
\else
\fi
\newcommand{\defeq}{:=}

\usepackage{color,soul,marginnote}

\ifarxiv
\title{Generative Bayesian Computation for Causal Inference}
\author{	
	\makebox[.4\linewidth]{Maria Nareklishvili}\\
	\textit{\small  Ragnar Frisch Centre for Economic Research}\\
	\textit{\small  University of Oslo}\\
	\and
	\makebox[.4\linewidth]{Nicholas Polson\footnote{Email: ngp@chicagobooth.edu}}\\
	\textit{\small  Booth School of Business}\\
	\textit{\small  University of Chicago}\\
	\and 
	\makebox[.4\linewidth]{Vadim Sokolov}\\
	\textit{\small  Department of Systems Engineering }\\
	\textit{\small  and Operations Research}\\
	\textit{\small  George Mason University}\\
}

\date{First Draft: June 20, 2023\\This Draft: December 20, 2024}
\else
\fi

\graphicspath{{./fig/}}

\begin{document}
\ifarxiv
\maketitle
\begin{abstract}
\noindent  Generative Bayesian Computation (GBC) methods are developed for Casual Inference. 
Generative methods are simulation-based methods that use a large
training dataset to represent posterior distributions as a map (a.k.a. optimal transport) to a base distribution. They avoid the use of  MCMC by replacing the conditional posterior inference problem with a supervised learning problem. We further propose the use Quantile ReLU networks which are density free and hence apply in a variety of Econometric settings where data generating processes  are specified by deterministic latent variables updates or as moment constraints. Generative approaches directly simulate large samples of observables and unobservable (parameters, latent variables) and then apply high-dimensional quantile regression to learn a nonlinear transport map from base distribution to parameter inference. We illustrate our methodology in the field of causal inference. Our approach can also handle nonlinearity and heterogeneity. Finally, we conclude with the directions for future research. 
\end{abstract}
\else
\fi

\newpage
\section{Introduction}\label{sec:intro}
Generative AI methods are proposed to solve problems of inference and prediction in econometrics. Generative methods require a use of large simulated training dataset which are prevalent in econometrics. The goal of such methods is to use deep neural networks to find a stochastic mapping between the parameters and data. Causal inference provides a natural testing grand for this methods. We develop NN architectures for this types of problems and future research is required for other problems such as DSGE models, auction models, IO models and others. There is a number of advantages over the simulation based techniques such as MCMC.  Our quantile gen-AI method avoids using densities and can be applied to high dimensions problems. The approach can be extended to solve decision-making problems using reinforcement learning methods, see \cite{polson2024generative}.

Our work builds on ideas from earlier papers that proposed using Bayesian non-linear models to analyze treatment effects and causality \cite{hill2007bayesian,hahn2020bayesian} and work on using deep learning for instrumented variables \cite{nareklishvili2022deep}. We study the implicit quantile neural networks \cite{polson2024generative}. Further, we investigate a long standing debate of causal inference on weather the propensity score is necessary for estimating the treatment effect. The intersection of the Bayesian methods, machine learning \citep{bhadra2021merging,xu2018bayesian} and causal inference in the context of observational data is a relatively new area of research. Both Bayesian and machine learning  techniques provide intuitive and flexible tools for analyzing complex data. Specifically, non-linearity's and heterogeneous effects can be modeled using both Bayes and ML techniques. Some authors propose a compromise between frequentist and Bayesian methods, for example, \cite{antonelli2022causal} consider using Bayesian methods to estimate both a propensity score and a response surface in the high-dimensional settings, and then using  a doubly-robust estimator by averaging over draws from the posterior distribution of the parameters of these models. \cite{stephens2023causal} argue that pure Bayesian methods are more suitable for causal inference.

Let $ ( X,Y) \sim P_{X,Y} $ be input-output pairs and $P_{X,Y} $ a joint measure. By factorizing this measure as $ P_X \times P_{Y|X} $ where $ P_{Y|X} $ is known as the forward map we can simulate a large  training dataset $ ( X_i , Y_i )_{i=1}^N \sim P_{X,Y} $. The goal is to characterize the inverse conditional map $ P_{X|Y} $. Standard prediction techniques  can be used to find the conditional posterior mean $ \hat{X} (Y) = E(X|Y) = f(Y) $ of the input given the output via a  multivariate  non-parametric  regression $ X = f(Y) + \epsilon $.   Typically estimators, $ \hat{f} $, include KNN and Kernel methods. Recently, deep learners have been proposed and the theoretical properties of superpositions of affine functions (a.k.a. ridge functions) have been provided (see \cite{diaconis1984nonlinear,montanelli2020error,schmidt-hieber2020nonparametric,polson2018posterior}).

Generative methods take this approach one step further. Let $ Z \sim P_Z $ be a base measure for a latent variable, $Z$, typically a standard multivariate normal or vector of uniforms. The goal of generative methods is to characterize the posterior measure $ P_{X|Y} $ from the training data
$ ( X_i , Y_i )_{i=1}^N \sim P_{X,Y} $ where $N$ is chosen to be suitably large. A deep learner is used to estimate $ \hat{G} $ via the non-parametric regression $ X = G(Y, Z ) $ . In the case, where $Z$ is a unfrock, this amounts to inverse cdf sampling, namely $ X = F_{X|Y}^{-1} ( U ) $. 
 
 \cite{polson2023generative} suggest the use of autoregressive RNNs for multivariate inference and \cite{kim2023deep} propose the use of Brenier’s maps as a solution to the multivariate case, see also \cite{carlier2016vector}.  One can therefore view Bayesian inference is then a high dimensional non-parametric regression in terms of a supervised learning problem.

Our goal then is to train a conditional deep learner t $ \hat{Q}_{Y|X} :  (U, \mathcal{X} )  \rightarrow \mathcal{Y} $ where  the conditional cdf is defined in terms of quantiles, $\tau$, via 
$$
\hat{Q}_{Y|X} ( \tau , x )  : \; ( \tau , x ) \rightarrow  \in \{ y \in \Re : F_{Y|X} ( y | x ) \geq \tau \} 
$$
This will allow us to simulate from the posterior using the inverse cdf method of von Neumann. The inherent difficulty in the multivariate case is that there exists multiple transport maps to determiner a generative model. One approach is to use an autoregressive RNN architecture or to use a Brenier’s map \cite{hutter2021minimax}. Specifically,  to learn an inverse CDF, we use a kernel trick known as cosine embedding and augment the predictor space. This has the effect of using. discrete cosine transform for $ \tau$. We use a different approach, to learn a single quantile function $F^{-1}_{Y|X} (\tau,x) = f_{\theta_f} (\tau,x)$, and then use the quantile function to generate samples from the target distribution. We represent the quantile function is a function of superposition for two other functions $F^{-1}(\tau,x) = f_{\theta_f}(\tau,x) = g(\psi(x)\circ \phi(\tau))$, as proposed in \cite{dabney2018implicit}, where $\circ$ is the element-wise multiplication operator. Both functions $g$ and $\phi$ are feed-forward neural networks. To avoid over-fitting, we use a sufficiently large training dataset. Another advantage of quantiles is that they naturally lead to estimates of posterior functionals such as means see \cite{polson2024generative}.  This is particularly useful in casual inference where one wants to learn the average treatment effect. 

\cite{polson2023generative} propose Generative Bayesian Computation (GBC) as an alternative to MCMC and ABC methods. They propose the use of quantile ReLU deep neural networks to avoid the need for densities.  

\begin{itemize}
    \item Theoretical results for multivariate quantile networks, see work on TS by \cite{gouttes2021probabilistic,ostrovski2018autoregressive,kronheim2021implicit}
    \item Relation between quantiles and Wasserstein distance. \cite{dabney2018implicit,ostrovski2018autoregressive}
    \item Gen AI is greater than GAN models. Explain the difference in objective functions. Why SGD has hard time optimizing the GAN objective function. Why quantile networks are easier to train. 
    \item Interpolation results for deep learning \cite{belkin2018understand,belkin2019does,belkin2019reconciling,telgarsky2015representation,telgarsky2016benefits,polson2018posterior,padilla2022quantile}
    \item Formal connection with ABC \cite{wang2023adversarial,kaji2022metropolis,kim2023deep,polson2023generative,nareklishvili2023generative}
\end{itemize}

\paragraph{Noise Outsourcing Theorem} If $ (X ,  Y ) $ are random variables in a Borel space $  ( \mathcal{X} , \mathcal{Y} ) $ then there exists an r.v. $U$ which is independent of
 $ Y$ and a function $ G^\star : [0,1] \times \mathcal{X} \rightarrow Y $ 
$$
(X , Y )  \stackrel{a.s.}{=} (X ,  G^\star ( U , X  ) )
$$
Hence the existence of $G^\star$  follows from the noise outsourcing theorem \cite{kallenberg1997foundations}. 

Moreover, if there is a conditional statistic $\hat{Z}(x) = E(Z| X= x ) $ with 
$ X \bot Y | \hat{Z} (X) $, then
$$
Y | X = x  , Z   \stackrel{a.s.}{=}  G^\star ( U , \hat{Z}(X)  )
$$
The role of $\hat{Z}(X)$ is a summary statistic. It performs dimension reduction in $n$, the dimensionality of the signal.   Here is replaces the allocation $Z$ by its
conditional mean 
$$ \hat{Z}( x) = E( Z | X=x ) = P ( Z=1 | X=x ) 
$$ 
This acts as a propensity score. The propensity score can be used to fill in some of the missing counterfactual values. 
Propensity Score is a summary statistic function $\pi(x) \rightarrow R$. A typical approach is to run a logistic regression of $Z$ on $x$.
\[
	\pi(x) = g\left( \hat{Z} (x) \right),
\]
where $g$ is a logit function. This approach guarantees that when the $x$'s from treatment group and $x$'s from control group are similar distributionally (histograms are similar) , the propensity scores are close. Other approaches include inverse weighting, stratification, matching, and subclassification.
An overview and importance of propensity score is given by \cite{rosenbaum1983central}, \cite{jiang2017} show that the optimal choice of summary (a.k.a. sufficient statistics \cite{kolmogorov1942definition}) is the conditional mean, Brillinger describes consistent procedure based on OLS to estimate $ \hat{Z} ( x ) $.  Much of the literature in casual inference has focused on running a logistic regression and using that in the
architecture to estimate the conditional distribution of the treatment effect $ \tau \defeq Y(1) - Y(0) $. In this theorem we will take $U \sim U(0,1) $ and the transport map $ G^\star$ to be the inverse cumulative distribution function $  F^{-1}_{  \bar{Y} | X = x }$.

A number of researchers have studied the theoretical properties of quantile deep learners in terms of their asymptotic properties.
This is relevant as we can choose $N$ the size of our training dataset. \cite{white1992nonparametric} uses the method of sieves to provide consistency results  for a single layer feed forward network. \cite{padilla2022quantile} provides minimax rates for $ \beta$-H\"older continuous functions.
\cite{schmidt-hieber2024generative} consider conditional quantile regression. \cite{carlier2016vector} and \cite{kim2023deep} consider the multivariate estimation of quantiles using a Brenier’s map. 

The rest of the paper is organized as follows. Section \ref{sec:lit} provides a brief overview of the existing literature. Section \ref{sec:method} describes the proposed method.  Section \ref{sec:results} presents the results of the simulation study. Section \ref{sec:conclusion} concludes.

\subsection{Causal AI}\label{sec:lit}

In studies of causality and treatment effects, each unit from $U$ (sample) has one of $k$ possible treatments. Thus a single treatment is assigned to each units. In a controlled experiment, the treatment is assigned randomly. However, we study the case of an observational data, when the treatment is not assigned randomly and the treatment effect may occur due to confounding (a.k.a. selection bias).  Selection bias is simply the dependency between the treatment assignment and the outcome $y$. The goal is to estimate the treatment effect. The treatment effect is defined as the difference between the outcome under treatment and the outcome under control. The outcome is a random variable $Y$ and the treatment is a random variable $Z$. The treatment effect is a random variable $\tau = Y(1) - Y(0)$. We assume that we observe the confounding predictors $x$, meaning that $x$ and $z$ are conditionally independent, given $x$. 

One way to approach the problem of estimating the treatment effect is to construct a counterfactual sample. The counterfactual sample is a hypothetical sample, where each unit has all possible treatments. More generally, the counterfactual process is then $Y(x,z)$ for all possible combinations of units $x$ and treatments $z$.  The realized sample (observed) on the other hand has only one observation per unit-treatment pair. In other words, the counterfactual process allows us to compute the conditional distribution of the response for the same unit under different treatments. The observed sample only allows us to compute the conditional distribution of the response for the same unit under the same treatment. One approach to causal inference is to estimate the counterfactual process from the observed sample. However, not everybody is enthusiastic about the approach of designing a counterfactual sample \cite{mccullagh2022ten}. For example, The \cite{dawid2000causal} argues, is that the counterfactual framework adds much to the vocabulary but brings nothing of substance to the conversation regarding observables. \cite{dawid2000causal} presents an alternative approach, based on Bayesian decision analysis. The main criticism is that there are multiple ways to construct the counterfactual samples and none of them are checkable.

In regression setting, the propensity score is a function of the conditional probability of treatment given the covariates. The propensity score is a sufficient statistic for the treatment assignment.  Then to estimate the treatment effect, we find ``similar'' units in the control group and compare their outcomes to the treatment group. The similarity is defined by the difference in propensity scores. In the controlled experiments the distributions over $X \mid Z=1$ and $X \mid Z=0$ should be the same. In observational studies, this is not the case.  The main difference between a traditional predictive model and the propensity score model $\pi(x)$ is that observed $y$'s are not used for training the propensity score model.
Furthermore, a common feature of the real-life problems is that the response function $f$ and the propensity score function $\pi$ are highly-non linear. Which makes many Bayesian methods inapplicable. For example, the propensity score matching is a popular method for causal inference. However, it is not clear how to apply this method in the Bayesian setting. The propensity score matching is a non-parametric method, which means that it does not require any assumptions about the functional form of the propensity score. However, the Bayesian approach requires a parametric model for the propensity score. Yet, another complicating factor can be deterministic relationships between the covariances and the treatment/outcome. In this case, sample-based Bayesian methods are not applicable.

Hahn et al (2020) propose the use of Bayesian casual forests as a direct extend of the BART sum-of-trees approach of \cite{hill2011bayesian}.
There are many other nonlinear regression (a.k.a supervised learning) approaches for estimation and inference of average treatment effects (ATE)
including double machine learning \cite{carlier2016vector} and generalised boosting \cite{mccaffrey2004propensity,mccaffrey2013tutorial}, More recently,  \cite{taddy2016nonparametric} focus on estimating heterogeneous effects from experimental rather than observational data. Conditional ATE (or CATE) using regression trees is proposed in \cite{wager2018estimation}. \cite{athey2019generalized} provide an inferential framework for CATE estimation. \cite{zaidi2018gaussian} use Gaussian processes and directly model the transomed response surface.

While some researchers \cite{banerjee2020road,duflo2007using} argue that randomized experiments can and should be used to estimate the treatment effect, it is the case that randomized experiments are not always possible and that observational studies can be used to estimate the treatment effect. \cite{rubin1974estimating} provides a good discussion of the difference in the estimation procedures for randomized and non-randomized studies.

It is contended that propensity score is not needed to estimate the treatment effects \cite{hahn1998role,hill2007bayesian}. On the other hand, \cite{rubin2006estimating} argues that estimating propensity score, it is hard to distinguish the treatment effect from the change-over-time effect. Another debate is wether Bayesian techniques or traditional frequentist approaches are more suitable for the econometrics applications \cite{stephens2023causal}.

The case of binary treatments \citep{splawa-neyman1990application} and propensity score approach have been thoroughly studied \cite{rubin1974estimating,holland1986statistics}. The counterfactual approach due to \cite{rubin1974estimating} is similar to the do-operator \cite{pearl2009causality}, in fact the two approaches are identical, when $Z$ is independent of $x$.  For Bayesian techniques see \cite{xu2018bayesian}. Machine learning techniques provide flexible approaches to more complex data generating processes, for example when networks are involved \cite{puelz2022graphtheoretic}. Tree based techniques are popular \cite{wager2018estimation}. For a  deep learning approach see \cite{vasilescu2022causal}.

\paragraph{Optimal rates of Statistical Learning.} Consider the non-parametric condition regression, $ y_i= f (x_i) + \epsilon_i $ where
$ x_i = ( x_{1i} , \ldots , x_{di} ) $. We wish to estimate a $d$-dimensional multivariate function $ f( x_1 , \ldots , x_d ) $ 
where $ x  = ( x_1 , \ldots , x_d ) \in [0,1]^d $. From a classical risk perspective, define
$$
R ( f , \hat{f}_N ) = E_{X,Y} \left ( \lVert  f - \hat{f}_N \rVert^2 \right ) 
$$
where $ \lVert . \rVert $ denotes $ L^2 ( P_X) $-norm.

Under standard assumptions, we have an optimal minimax rate $ \inf_{\hat{f}} \sup_f R( f , \hat{f}_N ) $ of
$ O_p \left ( N^{- 2 \beta /( 2 \beta + d )} \right ) $ for $\beta$-H\'older smooth functions $f$.
Typically, this space is too large as the bound still depends on $d$. By restricting the class of functions better rates can be obtained including ones that do no depend on $d$ and in this sense we avoid the curse of dimensionality.
For example, it is common to consider the class of linear superpositions (a.k.a. ridge functions) and projection pursuit. 

Another asymptotic result comes from a posterior concentration property. Here $ \hat{f}_N $ is constructed as a regularised  MAP (maximum a posteriori)
estimator which solves the optimisation problem
$$
\hat{f}_N = \arg \min_{ \hat{f}_N } \frac{1}{N} \sum_{i=1}^N ( y_i - \hat{f}_N ( x_i )^2 + \phi ( \hat{f}_N ) 
$$
The ensuing posterior distribution $ \Pi ( f  | x , y ) $ can be shown to have optimality results and concentrate on the minimax rate (up to a 
$ \log N $ factors).

A key result in the deep learning literature is the following rate. Given a training dataset of  input-output pairs $ ( x_i , y_i)_{i=1}^N $ from the model
$ y = f(x) + \epsilon $ where $f$ is a deep learner (a.k.a. superposition of functions), denoted by
$f = g_L \circ \ldots g_1 \circ g_0$,
where $g_i$ are $ \beta_i$-smooth H\'older functions with $ d_i $ variables, that is $ | g_i (x) -g_i (y) < | x-y |^{\beta_i} $.

Then the estimator has optimal rate
$$
O \left ( \max_{1\leq i \leq L } N^{- 2 \beta^* /( 2 \beta^* + d_i ) } \right )  \; {\rm where} \; \beta_i^* = \beta_i \prod_{l = i+1}^L \min ( \beta_l , 1 ) 
$$
Applying this to the class of generalised additive models 
$f_0 ( x ) = h \left ( \sum_{p=1}^d f_{0,p} (x_p) \right ) $
where 
$ g_0(z) = h(z) , \; g_1 ( x_1 , \ldots , x_d ) = ( f_{01}(x_1) , \ldots , f_{0d}(x_d) ) \; {\rm and} \;  g_2 ( y_1 , \ldots , y_d ) = \sum_{i=1}^d y_i  $.
So $ d_1 =1 , d_2 = 1 $ and $t_3 =1 $ as $h$ is Lipschitz.

Therefore, the optimal rate is $ O( N^{-1/3} ) $. Independent of $d$ which is the same as a deep ReLU network.
For $3$-times differentiable (cubic B-splines ), \cite{coppejans2004kolmogorovs} finds an optimal rate of $ O( N^{-3/7} ) = O( N^{-3/(2 \times 3 + 1) } ) $ matching the theory
developed above.

\section{Generative Bayesian Computation}\label{sec:method}
Let $Y$ denote a scalar response and $Z$ denote a binary treatment, and $x \in R^d$ be the covariates. We observe sample $(Y_i,Z_i,x_i)$, for $i=1,\ldots n$. We use $Y_i(0)$ and $Y_i(1)$ to denote the outcome (hypothetical) with treatment zero or one. The observed outcome is given by
\[
	Y_i = Y_i(0) + Z_i(Y_i(1)-Y_i(0)).
\] 
We assume that the outcome is conditionally independent of the assigned treatment given the covariates, i.e., $Y_i(0)$ and $Y_i(1)$ are independent of $Z_i$ given $x_i$. We also assume that 
\[
	P(Z_i=1\mid x_i) >0.
\]
The first condition assumes we have no unmeasured confounders. Given the two assumptions above, we can write the conditional mean of the outcome as
\[
	\tau(x_i) = E[Y_i(1)-Y_i(0)\mid x_i].
\]
The goal is build a predictive model 
\[
	Y_i = f(x_i,Z_i, \pi(x_i)) + \epsilon_i,~ \epsilon_i \sim N(0,\sigma^2_i),
\]
where $\pi(x_i)$ is the propensity score function. Then
\[
	\tau(x_i) = f(x_i,1) - f(x_i,0).
\]
Casual inference then can be viewed as a missing data problem.The "complete data" is the bivariate potential outcomes $ ( Y(1) , Y()) ) $. We only get to
see one of these at a time.
Let $Y$ denote a scalar response, $Z$ a binary treatment indicator. Let $x$ denote a $d$-dimensional vector of observed control variables. 
Consider an observed sample of size $n$, denoted by
$$
(Y_i , Z_i , x_i ) , 1 \leq i \leq n .
$$
We are interested in estimating various treatment effects. Conditional ATE effects correspond to  $E(Y_i | X_i , Z_i=1 ) $ and $E(Y_i | X_i , Z_i=0 ) $.
We observe the potential outcome that corresponds to the realized treatment
$$
Y_i = Z_i Y_i(1) + ( 1 - Z_i ) Y_i (0) .
$$
Throughout we will make  four assumptions that we make for casual estimation to be valid.
In particular, strong ignobility and positivity given by 

\begin{itemize}
\item Consistency:  Observed data is unrelated to the potential outcomes via the identity
$$
Y = Y(1) Z + Y(0) ( 1 - Z ) 
$$
\item Non-inference:  For any samples size, 
$$ 
( Y_j(1), Y_j(0) \bot Z_j ) \; \; \forall  \; i , j \in \{ 1 ,\ldots , n \} 
$$ 
\item Positivity:
$0 < P( Z=1 | X= x ) < 1 \; \forall x \in \mathcal{X} $.
\item Conditional Unconfoundedness:
The following conditional independence condition holds
$$
( Y(1), Y(0) ) \bot Z_j  | X 
$$
\end{itemize}
In frequentist approaches, adjustment is conducted by estimating parameters independently in the propensity score model $\pi$ and the outcome model $f, \epsilon$. However, this two-step analysis is leads to inefficiencies. Instead, it is more intuitive to develop a single joint model that encompasses both the treatment and outcome variables. As a result, there has been a discussion regarding the applicability of Bayesian methods in causal analysis. The literature on advanced techniques for conducting Bayesian causal analysis is expanding, but certain aspects of these methodologies appear unconventional.

\subsection{GBC for Casual Inference}

Hence we have $ \tau = Y(1) - Y(0) $ and. $ \tau(x) \defeq E \left ( \tau |  X = x \right  ) = E ( Y(1) - Y(0) | X=x )  $ be  the average treatment effect ATE . Hence, we can write
$$
\tau (x_i) = E(Y_i | X_i , Z_i=1 ) - E(Y_i | X_i , Z_i=0 ) 
$$
From a modeling perspective, 
$Y_i = f( x _i , Z_i ) + \epsilon_i ,\; \;  \epsilon_i \sim N(0, \sigma^2 )  $
with  $ E(Y_i | X_i , Z_i =z_i) = f(x_i ,z_i) $ where $ Z_i =0 , 1 $.
Alternatively, Hahn et al recommend repressing the response surface as
$$
 E(Y_i | X_i , Z_i =z_i) = \mu ( x_i , \hat{\pi} (x_i) ) + \tau(x_i) z_i 
$$
where $ \mu$ and $ \tau $ are given independent BART priors.

We want quantiles  to calculate the treatment effect, namely the function $ F^{-1}_{ \tau | x } ( \cdot ) $.
 We start by simulating $\theta_i, y_i$ pairs from prior and the forward model, then we reverse 
\[
	\pi(x_i) = P(Z_i=1\mid x_i) = E(Z_i\mid x_i)
\]
So we have sufficient statistics and can replace the dataset with $\pi(x_i),y_i$. We can then use the quantile regression to estimate the quantiles of $Y_i(1)-Y_i(0)$ given $x_i$.
\[
	p(y,z\mid x) = p(y\mid x,z)p(z\mid x)
\]
\[
	y(x,1) = y(x,0) = H(y,x,\pi(x)),~\pi(x) = E(Z\mid x)
\]
Having fitted the deep neural network, we can use the estimated inverse map to evaluate at new $y$ and $\tau$ to obtain a set of posterior samples for any new $y$. The caveat being is to how to choose $N$ and how well the deep neural network interpolates for the new inputs. We also have flexibility in choosing the distribution of $\tau$, for example, we can also for $\tau$ to be a high-dimensional vector of Gaussians, and essentially provide a mixture-Gaussian approximation for the set of posterior. MCMC, in comparison, is computationally expensive and needs to be re-run for any new data point. 

The idea is  quite simple. Use non-parametric regression to estimate $ f(x_i  , 0) $ and $ f(x_i ,1) $
Our approach differs from traditional approaches  in that we use conditional quantile function estimators based on deep ReLU networks.
CATE is then estimated as an integrated quantile function by appealing to the following quantile identity for expectation given by 
$$
\hat{\tau} ( x ) = E \left ( \tau | X = x \right  )  = \int_0^1 F^{-1}_{  \bar{Y} | X = x } ( U ) d U 
$$
Hence, we need only find a distributional generator for the conditional distribution posterior of $ p  ( \bar{Y} | X= x ) $.

GBC in a simple way is using pattern matching to provide a look-up table for the map from data to treatment effect. Bayesian computation has then being replaced by the optimisation performed by Stochastic Gradient Descent (SGD). In our examples, we discuss choices of architectures for $H$ and $S$. Specifically, we propose cosine-embedding for transforming $\tau$. 

Hence, our addition to the literature is the use of generative methods which take advantage of the implicit model for the counterfactual response.
Moreover, the use of deep ReLU networks (a.k,a. hyperplanes) is more flexible than trees (a.k.a. cylinder sets).
Quantile ReLU estimator.

\vspace{0.1in}



\paragraph{Double Descent} There is still the question of approximation and the interpolation properties of a DNN. Recent research on interpolation properties of quantile neural networks were recently studied by \cite{padilla2022quantile} and \cite{shen2021deep}, \cite{schmidt-hieber2020nonparametric}. See also \cite{bach2024high,belkin2019reconciling}
Shallow Deep Learners are known to provide good representations of multivariate functions and are good interpolators.

Hence even if $ y_{\mathrm{obs}}  $ is not in the simulated input-output dataset $ y_N $  we can still learn the posterior map of interest.
The Kolmogorov-Arnold theorem says any multivariate function can be expressed this way.  So in principle if $N $ is large enough we can learn the manifold structure in the parameters for any arbitrary nonlinearity. As the dimension of the data $y$ is large, in practice, this requires providing an efficient architecture. The main  question of interest. We recommend quantile neural networks. RelU and tanh networks are also natural candidates.

\paragraph{Deep Learning for Propensity Sceores}

The usual modeling approach is to use logistic regression. An alternative is to use a deep learner.
\cite{jiang2017} proposes the following architecture for the summary statistic neural network
\begin{align*}
	H^{(1)} = & ReLU \left(W^{(0)}H^{(0)}+b^{(0)}\right)\\
	H^{(2)} = & ReLU \left(W^{(1)}H^{(1)}+b^{(1)}\right)\\
	& \vdots\\
	H^{(L)} = & ReLU \left(W^{(L-1)}H^{(L-1)}+b^{(L-1)}\right)\\
	\hat\pi(x) = & W^{(L)}H^{(L)}+b^{(L)},
\end{align*}
where $H^{(0)} = Z$ is the input, and $\hat Z$ is the summary statistic output. 

The following algorithms  summarize our approach 

\begin{algorithm}
   \caption{Gen-AI for Bayesian Computation (GenAI-Bayes)}\label{algorithm_a}
\begin{algorithmic}[Gen-AI-Bayes]
   \STATE Simulate $\theta^{(i)} \sim p(\theta)$. Simulate $y^{(i)} \mid \theta^{(i)} \sim p(y\mid \theta)$, $i=1,\ldots,N$ or $ y^{(i)} = s( \theta^{(i)} ) $.
   \STATE Train $H$ and $S$, using $\theta^{(i)} = H(S(y^{(i)},\epsilon^{(i)})$, where $\epsilon^{(i)} \sim N(0,\sigma_{\epsilon})$
   \STATE For a given $y$, calculate a sample from $p(\theta \mid y)$ by $\theta \stackrel{D}{=} H(y , \tau )$ where  $\tau \sim U(0,1 )$
\end{algorithmic}
\end{algorithm}

\section{Application}\label{sec:results}

\subsection{Simulated Example} 

In this section we provide empirical examples and compare our approach with various alternatives. Specifically, we compare our method with generalized random forests \cite{athey2019generalized,wager2018estimation} and more traditional propensity score-based methods \cite{imbens2015causal}. Our synthetic data in generated using heterogeneous treatment effects and nonlinear conditional expectation function (response surface) and a sample size $n$ of 1000. We use a three-dimensional $(p=3)$ covariate with all three components drawn from standard normal distribution 
The data generating process is given by 
\begin{align*}
	x_{ij} \sim & N(0,1),~x \in R^{n\times p}\\
	\mu_i = & -6 + I(x_{i1}> x_{i2}) + 6|x_{i2}-1| \mbox{ (Nonlinear effect)}\\
	\pi_i \sim & \sigma(\mu_i) ~\mbox{ (Sigmoid)}\\ 
	z_i \sim & \mathrm{Bernoulli}(\pi_i)\\
	\tau_i = & 1-2x_{i2}x_{i3} \mbox{ (Nonlinear treatment effect)}\\
	E(y_i | x_i) \sim & \mu_i + \tau_i z_i\\
	y_i  | x_i \sim & N(E(y_i | x_i),\sigma^2)
 \end{align*}
 Figure \ref{fig:hist-synthetic} below shows the histograms of generated $y$, $\mu$, and $\tau$. Notice, that we standardized $\tau$ to be of mean zero and variance of one. 
\begin{figure}[H]\label{fig:hist-synthetic}
	\includegraphics[width=1\textwidth]{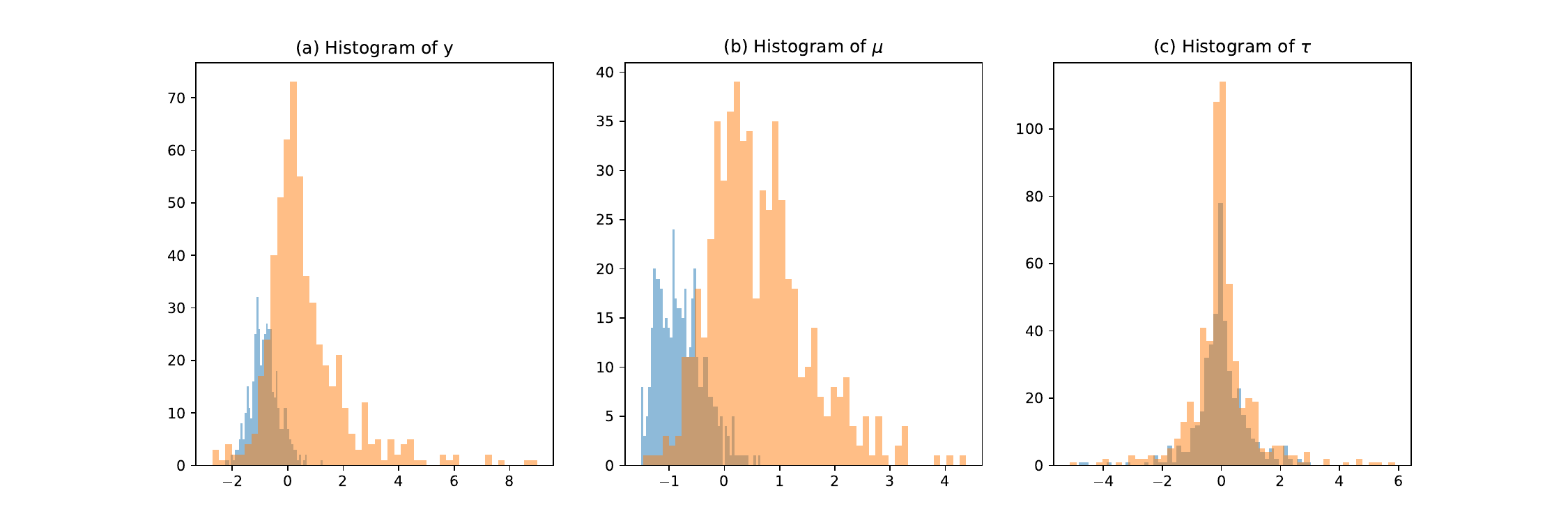}
	\caption{Synthetic data histograms}
\end{figure}
We calculate three metrics to evaluate and benchmark our method. We consider the average treatment effect (ATE) calculated from the sample and compute mean squared error (MSE) as well as coverage and average interval length. Further, we consider conditional average treatment effect (CATE), averaged over the sample.

\begin{figure}[H]\label{fig:fit-synthetic}
	\includegraphics[width=1\textwidth]{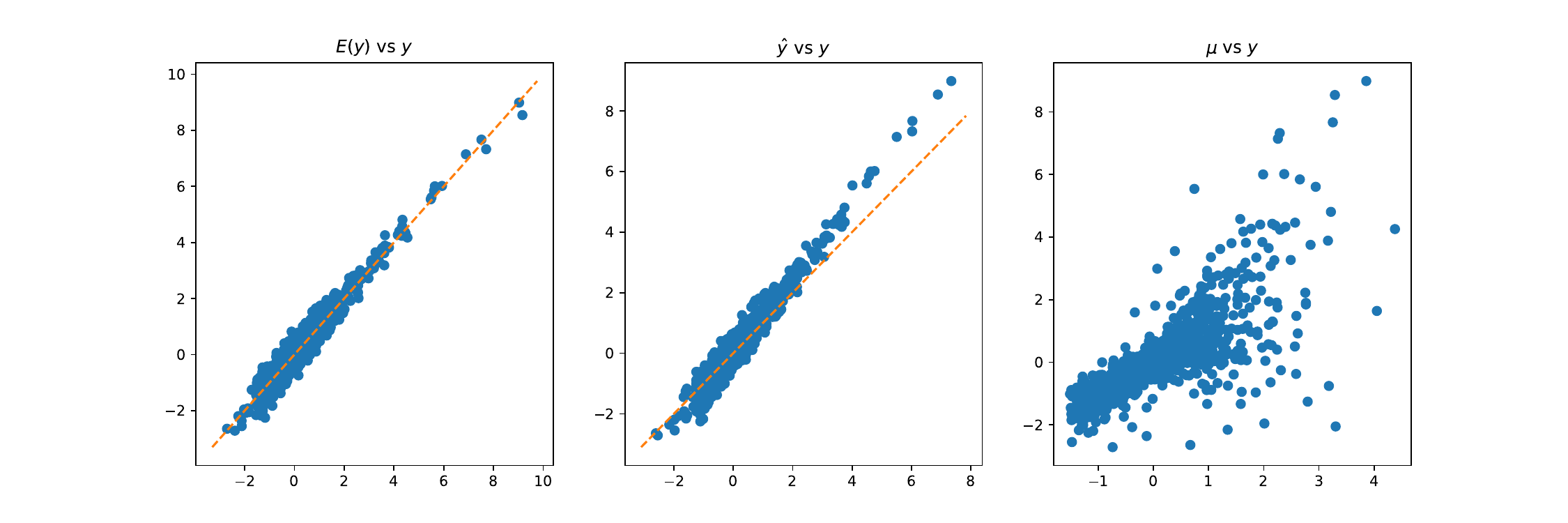}
	\caption{The middle plot compares fitted reposes $\hat y$ and simulated ones $y$. Left plot compares the simulated $y$ vs the noiseless values $E(y)$. Right plot shows $y$ vs $\mu$.}
\end{figure}

First, we show some plots that demonstrate the quality of our fit of the response, shown in Figure \ref{fig:fit-synthetic}.

We use neural networks as building blocks of our model. Each layer of a neural network is a function of the form
\[
f(x ; W, l) = h(Wx), ~ x\in \Re^d,~ W\in \Re^{d\times l},
\]
where $h$ is a nonlinear  univariate function, such as ReLU, applied element-wise to $x$, and $l$ is the number of neurons in the layer. We use the following architecture for the response surface. We start by calculating a cosine embedding of of the quantile $q$
\[
	s = ( \cos(i\pi q),  \ldots ,  \cos( 32 \pi q ) ) 
\]
Representing a discrete cosine transform embedding. The kernel embedding trick allows us to identify the density more efficiently with this set of basis functions. 
We use the following architecture: 
\begin{align*}
	s = & f(s; W_1,32)\\
	\tilde \pi = & f(x; W_2,8)\\ 
	\hat \pi = & f(\tilde \pi; W_3,32)\\
    \hat z =  &\sigma(\hat \pi)\\
	\mu = &  q \circ f([x,\tilde \pi]; W_3,32)\\
	\tau = & q \circ f(x; W_5,32)\\
	\hat y = &W_6(\mu + \tau \circ \hat z), ~W_6\in \Re^{32\times 2}.
\end{align*}
Here $\circ$ stands for element-wise multiplication. Our model generates a two-dimensional output $\hat y$, first element is the mean response and the second is the quantile response. We use the following loss function to jointly estimate the components of our model 
\begin{align*}
	q \sim & U(0,1)\\
	l_z =& (1/n)\sum_{i=1}^{n}z_i \log \hat z_i + (1-z_i)\log(1-\hat z_i)\\
	e_i =& y_i - \hat y_i\\
	l_{\mathrm{MSE}} = & (1/n)\sum_{i=1}^{n} e_{i1}^2\\
	l_q = & (1/n)\sum_{i=1}^{n} \max(q e_{i2},(q-1)e_{i2})\\
	l = & w_1 l_z + w_2 l_q + w_3 l_{\mathrm{MSE}}\\
\end{align*}
We add a constraint to the loss function to prevent the quantiles to cross, specifically our constraints are
\[
\begin{cases}
\hat y(\tau) < y,  & \mbox{ when } \tau < 0.5 \\
\hat y(\tau) > y,  & \mbox{ when } \tau > 0.5.
\end{cases}
\]
We add this constraint as a penalty term to the loss function.

\begin{figure}[H]
	\includegraphics[width=1\textwidth]{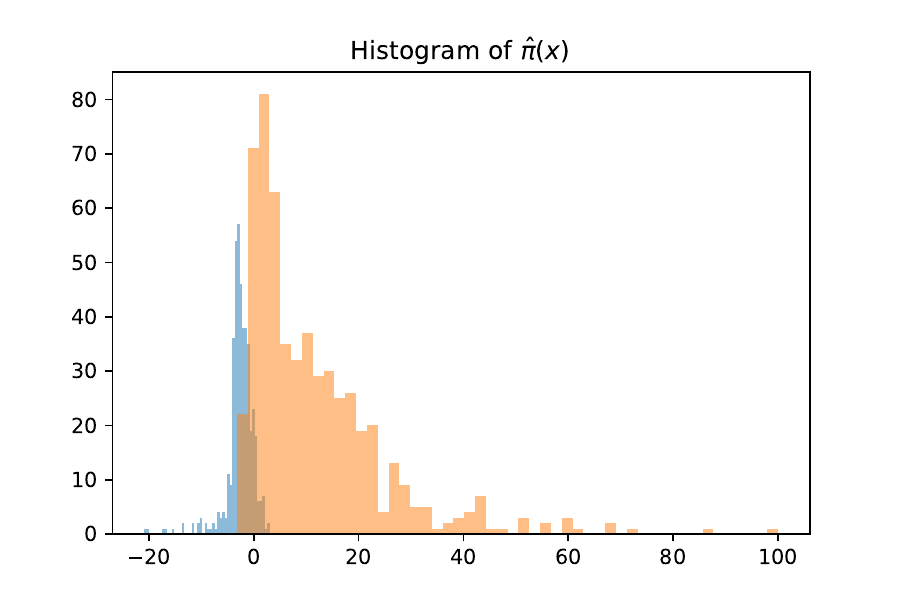}
	\caption{Histogram of fitted propensity scores $\hat \pi(x)$}
\end{figure}

\begin{figure}[H]\label{fig:tau-hist}
\begin{tabular}{cc}
	\includegraphics[width=0.5\textwidth]{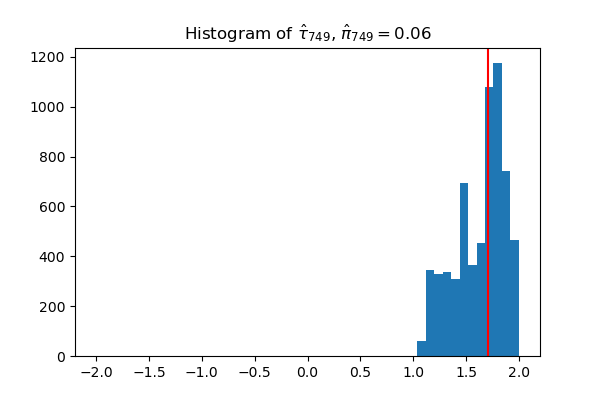} & \includegraphics[width=0.5\textwidth]{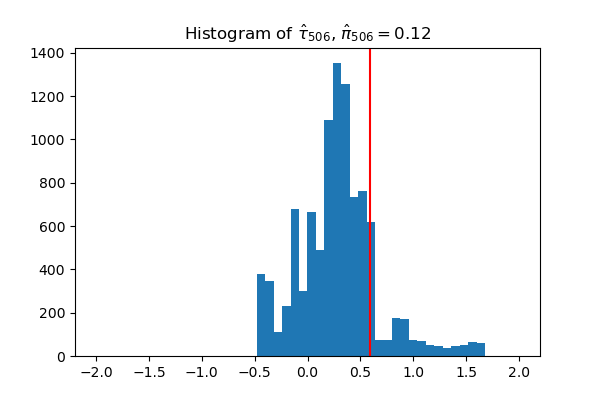}\\
	\includegraphics[width=0.5\textwidth]{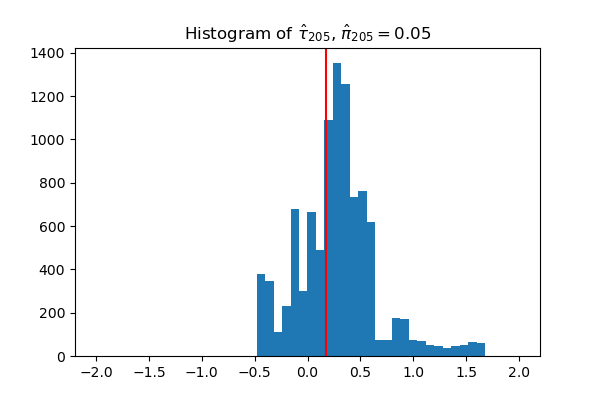} & \includegraphics[width=0.5\textwidth]{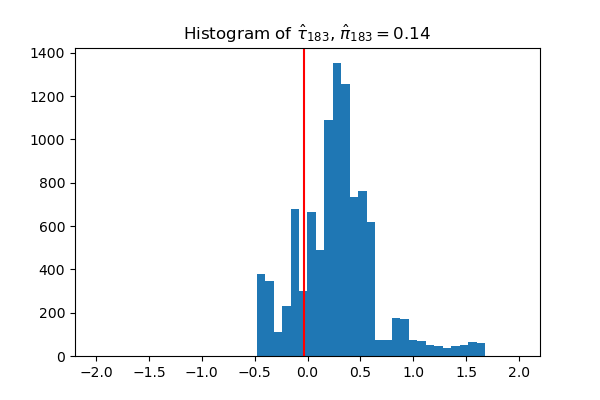}\\
	\includegraphics[width=0.5\textwidth]{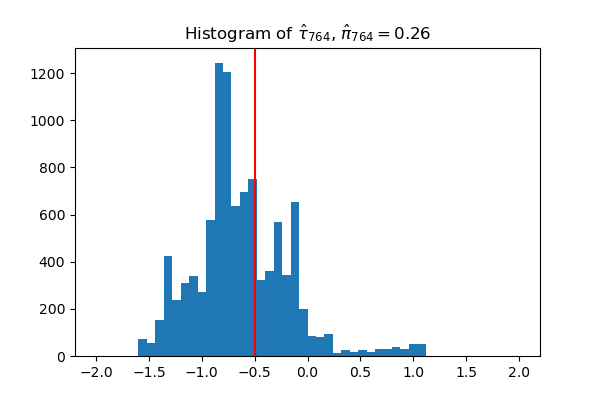} & \includegraphics[width=0.5\textwidth]{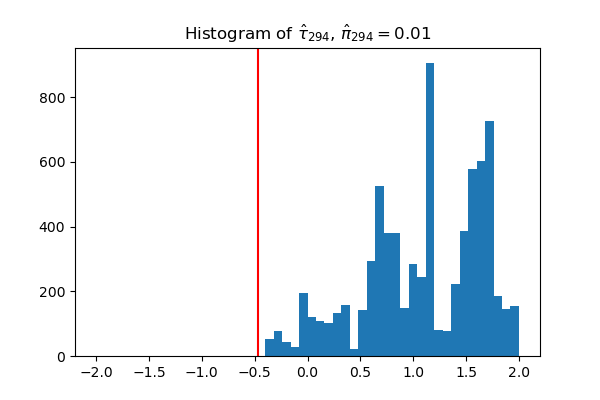}\\
\end{tabular}
\caption{Histogram of posterior values of treatment effect $\tau$ for randomly selected units that were assigned no treatment ($z=0$).}
\end{figure}

Figure \ref{fig:tau-hist} shows the posterior distribution of the treatment effect $\tau$ for randomly selected units that were assigned no treatment ($z=0$). The vertical red line is the true value of the treatment effect. The posterior distribution of $\tau$ is also very tight, which is consistent with the fact that the control group is large.

\section{Discussion}\label{sec:conclusion}
Generative methods differ from traditional simulation based tools in that they use a large training data set to infer predictive mappings rather than density methods The main tool is high-dimensional nonlinear nonparametric regression using deep neural networks. Inference for the observed data is then evaluation of the network and is therefore an interpolation approach to inference. There are many avenues for future research. Given wide applicability of simulation in econometrics models, designing architectures for specific problems is a a paramount interest. 

\nocite{*}
\bibliography{ref,Econ}

\begin{thebibliography}{71}
\providecommand{\natexlab}[1]{#1}
\providecommand{\url}[1]{\texttt{#1}}
\expandafter\ifx\csname urlstyle\endcsname\relax
  \providecommand{\doi}[1]{doi: #1}\else
  \providecommand{\doi}{doi: \begingroup \urlstyle{rm}\Url}\fi

\bibitem[Antonelli et~al.(2022)Antonelli, Papadogeorgou, and Dominici]{antonelli2022causal}
Joseph Antonelli, Georgia Papadogeorgou, and Francesca Dominici.
\newblock Causal inference in high dimensions: {{A}} marriage between {{Bayesian}} modeling and good frequentist properties.
\newblock \emph{Biometrics}, 78\penalty0 (1):\penalty0 100--114, March 2022.

\bibitem[Athey et~al.(2019)Athey, Tibshirani, and Wager]{athey2019generalized}
Susan Athey, Julie Tibshirani, and Stefan Wager.
\newblock Generalized random forests.
\newblock \emph{The Annals of Statistics}, 47\penalty0 (2):\penalty0 1148--1178, April 2019.

\bibitem[Bach(2024)]{bach2024high}
Francis Bach.
\newblock High-dimensional analysis of double descent for linear regression with random projections.
\newblock \emph{SIAM Journal on Mathematics of Data Science}, 6\penalty0 (1):\penalty0 26--50, 2024.

\bibitem[Banerjee et~al.(2020)Banerjee, Duflo, and Qian]{banerjee2020road}
Abhijit Banerjee, Esther Duflo, and Nancy Qian.
\newblock On the road: {{Access}} to transportation infrastructure and economic growth in {{China}}.
\newblock \emph{Journal of Development Economics}, 145:\penalty0 102442, 2020.

\bibitem[Belkin et~al.(2018)Belkin, Ma, and Mandal]{belkin2018understand}
Mikhail Belkin, Siyuan Ma, and Soumik Mandal.
\newblock To {{Understand Deep Learning We Need}} to {{Understand Kernel Learning}}.
\newblock In \emph{Proceedings of the 35th {{International Conference}} on {{Machine Learning}}}, pages 541--549. PMLR, July 2018.

\bibitem[Belkin et~al.(2019{\natexlab{a}})Belkin, Hsu, Ma, and Mandal]{belkin2019reconciling}
Mikhail Belkin, Daniel Hsu, Siyuan Ma, and Soumik Mandal.
\newblock Reconciling modern machine-learning practice and the classical bias--variance trade-off.
\newblock \emph{Proceedings of the National Academy of Sciences}, 116\penalty0 (32):\penalty0 15849--15854, August 2019{\natexlab{a}}.

\bibitem[Belkin et~al.(2019{\natexlab{b}})Belkin, Rakhlin, and Tsybakov]{belkin2019does}
Mikhail Belkin, Alexander Rakhlin, and Alexandre~B. Tsybakov.
\newblock Does data interpolation contradict statistical optimality?
\newblock In \emph{Proceedings of the {{Twenty-Second International Conference}} on {{Artificial Intelligence}} and {{Statistics}}}, pages 1611--1619. PMLR, April 2019{\natexlab{b}}.

\bibitem[Bernardo(1979)]{bernardo1979expected}
Jose~M. Bernardo.
\newblock Expected {{Information}} as {{Expected Utility}}.
\newblock \emph{The Annals of Statistics}, 7\penalty0 (3), May 1979.

\bibitem[Bhadra et~al.(2021)Bhadra, Datta, Polson, Sokolov, and Xu]{bhadra2021merging}
Anindya Bhadra, Jyotishka Datta, Nick Polson, Vadim Sokolov, and Jianeng Xu.
\newblock Merging two cultures: deep and statistical learning.
\newblock \emph{arXiv preprint arXiv:2110.11561}, 2021.

\bibitem[Carlier et~al.(2016)Carlier, Chernozhukov, and Galichon]{carlier2016vector}
Guillaume Carlier, Victor Chernozhukov, and Alfred Galichon.
\newblock Vector quantile regression: {{An}} optimal transport approach.
\newblock \emph{The Annals of Statistics}, 44\penalty0 (3):\penalty0 1165--1192, June 2016.

\bibitem[Coppejans(2004)]{coppejans2004kolmogorovs}
Mark Coppejans.
\newblock On {{Kolmogorov}}'s representation of functions of several variables by functions of one variable.
\newblock \emph{Journal of Econometrics}, 123\penalty0 (1):\penalty0 1--31, November 2004.

\bibitem[Dabney et~al.(2018)Dabney, Ostrovski, Silver, and Munos]{dabney2018implicit}
Will Dabney, Georg Ostrovski, David Silver, and R{\'e}mi Munos.
\newblock Implicit {{Quantile Networks}} for {{Distributional Reinforcement Learning}}, June 2018.

\bibitem[Dawid(2000)]{dawid2000causal}
A.~P. Dawid.
\newblock Causal {{Inference}} without {{Counterfactuals}}.
\newblock \emph{Journal of the American Statistical Association}, 95\penalty0 (450):\penalty0 407--424, June 2000.

\bibitem[Diaconis and Shahshahani(1984)]{diaconis1984nonlinear}
Persi Diaconis and Mehrdad Shahshahani.
\newblock On nonlinear functions of linear combinations.
\newblock \emph{SIAM Journal on Scientific and Statistical Computing}, 5\penalty0 (1):\penalty0 175--191, 1984.

\bibitem[Duflo et~al.(2007)Duflo, Glennerster, and Kremer]{duflo2007using}
Esther Duflo, Rachel Glennerster, and Michael Kremer.
\newblock Using {{Randomization}} in {{Development Economics Research}}: {{A Toolkit}}.
\newblock In T.~Paul Schultz and John~A. Strauss, editors, \emph{Handbook of {{Development Economics}}}, volume~4, pages 3895--3962. Elsevier, January 2007.

\bibitem[Gallant and White(1988)]{gallant1988there}
A.~Ronald Gallant and Halbert White.
\newblock There exists a neural network that does not make avoidable mistakes.
\newblock In \emph{Proceedings of the {{Second Annual IEEE Conference}} on {{Neural Networks}}, {{San Diego}}, {{CA}}, {{I}}}, 1988.

\bibitem[Gouttes et~al.(2021)Gouttes, Rasul, Koren, Stephan, and Naghibi]{gouttes2021probabilistic}
Ad{\`e}le Gouttes, Kashif Rasul, Mateusz Koren, Johannes Stephan, and Tofigh Naghibi.
\newblock Probabilistic {{Time Series Forecasting}} with {{Implicit Quantile Networks}}, July 2021.

\bibitem[Hahn(1998)]{hahn1998role}
Jinyong Hahn.
\newblock On the {{Role}} of the {{Propensity Score}} in {{Efficient Semiparametric Estimation}} of {{Average Treatment Effects}}.
\newblock \emph{Econometrica}, 66\penalty0 (2):\penalty0 315--331, 1998.

\bibitem[Hahn et~al.(2018)Hahn, Carvalho, Puelz, and He]{hahn2018regularization}
P.~Richard Hahn, Carlos~M. Carvalho, David Puelz, and Jingyu He.
\newblock Regularization and {{Confounding}} in {{Linear Regression}} for {{Treatment Effect Estimation}}.
\newblock \emph{Bayesian Analysis}, 13\penalty0 (1):\penalty0 163--182, March 2018.

\bibitem[Hahn et~al.(2020)Hahn, Murray, and Carvalho]{hahn2020bayesian}
P.~Richard Hahn, Jared~S. Murray, and Carlos~M. Carvalho.
\newblock Bayesian {{Regression Tree Models}} for {{Causal Inference}}: {{Regularization}}, {{Confounding}}, and {{Heterogeneous Effects}} (with {{Discussion}}).
\newblock \emph{Bayesian Analysis}, 15\penalty0 (3):\penalty0 965--1056, September 2020.

\bibitem[Hill and McCulloch(2007)]{hill2007bayesian}
Jennifer Hill and Robert McCulloch.
\newblock Bayesian {{Nonparametric Modeling}} for {{Causal Inference}}.
\newblock Technical report, June 2007.

\bibitem[Hill(2011)]{hill2011bayesian}
Jennifer~L. Hill.
\newblock Bayesian {{Nonparametric Modeling}} for {{Causal Inference}}.
\newblock \emph{Journal of Computational and Graphical Statistics}, 20\penalty0 (1):\penalty0 217--240, January 2011.

\bibitem[Hirano et~al.(2003)Hirano, Imbens, and Ridder]{hirano2003efficient}
Keisuke Hirano, Guido~W. Imbens, and Geert Ridder.
\newblock Efficient {{Estimation}} of {{Average Treatment Effects Using}} the {{Estimated Propensity Score}}.
\newblock \emph{Econometrica}, 71\penalty0 (4):\penalty0 1161--1189, 2003.

\bibitem[Holland(1986)]{holland1986statistics}
Paul~W. Holland.
\newblock Statistics and {{Causal Inference}}.
\newblock \emph{Journal of the American Statistical Association}, 81\penalty0 (396):\penalty0 945--960, 1986.

\bibitem[H{\"u}tter and Rigollet(2021)]{hutter2021minimax}
Jan-Christian H{\"u}tter and Philippe Rigollet.
\newblock Minimax estimation of smooth optimal transport maps.
\newblock \emph{The Annals of Statistics}, 49\penalty0 (2):\penalty0 1166--1194, April 2021.

\bibitem[Imbens and Rubin(2015)]{imbens2015causal}
Guido~W Imbens and Donald~B Rubin.
\newblock \emph{Causal Inference in Statistics, Social, and Biomedical Sciences}.
\newblock Cambridge University Press, 2015.

\bibitem[Jiang et~al.(2017)Jiang, Wu, Zheng, and Wong]{jiang2017}
Bai Jiang, Tung-Yu Wu, Charles Zheng, and Wing~H. Wong.
\newblock Learning {{Summary Statistic For Approximate Bayesian Computation Via Deep Neural Network}}.
\newblock \emph{Statistica Sinica}, 27\penalty0 (4):\penalty0 1595--1618, 2017.

\bibitem[Kaji and Ro{\v c}kov{\'a}(2022)]{kaji2022metropolis}
Tetsuya Kaji and Veronika Ro{\v c}kov{\'a}.
\newblock Metropolis--{{Hastings}} via {{Classification}}.
\newblock \emph{Journal of the American Statistical Association}, pages 1--15, May 2022.

\bibitem[Kallenberg(1997)]{kallenberg1997foundations}
Olav Kallenberg.
\newblock \emph{Foundations of {{Modern Probability}}}.
\newblock Springer, 2nd ed. edition edition, January 1997.
\newblock ISBN 978-0-387-94957-4.

\bibitem[Kim and Rockova(2023)]{kim2023deep}
Jungeum Kim and Veronika Rockova.
\newblock Deep {{Bayes Factors}}, December 2023.

\bibitem[Kolmogorov(1942)]{kolmogorov1942definition}
{\relax AN}~Kolmogorov.
\newblock Definition of center of dispersion and measure of accuracy from a finite number of observations (in {{Russian}}).
\newblock \emph{Izv. Akad. Nauk SSSR Ser. Mat.}, 6:\penalty0 3--32, 1942.

\bibitem[Kronheim et~al.(2021)Kronheim, Kuchera, Prosper, and Ramanujan]{kronheim2021implicit}
Braden Kronheim, Michelle~P. Kuchera, Harrison~B. Prosper, and Raghuram Ramanujan.
\newblock Implicit {{Quantile Neural Networks}} for {{Jet Simulation}} and {{Correction}}, November 2021.

\bibitem[Lasserre et~al.(2006)Lasserre, Bishop, and Minka]{lasserre2006principled}
Julia~A Lasserre, Christopher~M Bishop, and Thomas~P Minka.
\newblock Principled hybrids of generative and discriminative models.
\newblock In \emph{2006 IEEE Computer Society Conference on Computer Vision and Pattern Recognition (CVPR'06)}, volume~1, pages 87--94. IEEE, 2006.

\bibitem[McCaffrey et~al.(2004)McCaffrey, Ridgeway, and Morral]{mccaffrey2004propensity}
Daniel~F. McCaffrey, Greg Ridgeway, and Andrew~R. Morral.
\newblock Propensity {{Score Estimation With Boosted Regression}} for {{Evaluating Causal Effects}} in {{Observational Studies}}.
\newblock \emph{Psychological Methods}, 9\penalty0 (4):\penalty0 403--425, 2004.

\bibitem[McCaffrey et~al.(2013)McCaffrey, Griffin, Almirall, Slaughter, Ramchand, and Burgette]{mccaffrey2013tutorial}
Daniel~F. McCaffrey, Beth~Ann Griffin, Daniel Almirall, Mary~Ellen Slaughter, Rajeev Ramchand, and Lane~F. Burgette.
\newblock A tutorial on propensity score estimation for multiple treatments using generalized boosted models.
\newblock \emph{Statistics in Medicine}, 32\penalty0 (19):\penalty0 3388--3414, 2013.

\bibitem[McCullagh(2022)]{mccullagh2022ten}
Peter McCullagh.
\newblock \emph{Ten {{Projects}} in {{Applied Statistics}}}.
\newblock Springer {{Series}} in {{Statistics}}. Springer International Publishing, Cham, 2022.
\newblock ISBN 978-3-031-14274-1 978-3-031-14275-8.

\bibitem[Montanelli and Yang(2020)]{montanelli2020error}
Hadrien Montanelli and Haizhao Yang.
\newblock Error bounds for deep {{ReLU}} networks using the {{Kolmogorov--Arnold}} superposition theorem, May 2020.

\bibitem[Nareklishvili et~al.(2022{\natexlab{a}})Nareklishvili, Polson, and Sokolov]{nareklishvili2022deep}
Maria Nareklishvili, Nicholas Polson, and Vadim Sokolov.
\newblock Deep partial least squares for iv regression.
\newblock \emph{arXiv preprint arXiv:2207.02612}, 2022{\natexlab{a}}.

\bibitem[Nareklishvili et~al.(2022{\natexlab{b}})Nareklishvili, Polson, and Sokolov]{nareklishvili2022feature}
Maria Nareklishvili, Nicholas Polson, and Vadim Sokolov.
\newblock Feature selection for personalized policy analysis.
\newblock \emph{arXiv preprint arXiv:2301.00251}, 2022{\natexlab{b}}.

\bibitem[Nareklishvili et~al.(2023{\natexlab{a}})Nareklishvili, Polson, and Sokolov]{nareklishvili2023deep}
Maria Nareklishvili, Nicholas Polson, and Vadim Sokolov.
\newblock Deep partial least squares for instrumental variable regression.
\newblock \emph{Applied Stochastic Models in Business and Industry}, 2023{\natexlab{a}}.

\bibitem[Nareklishvili et~al.(2023{\natexlab{b}})Nareklishvili, Polson, and Sokolov]{nareklishvili2023generative}
Maria Nareklishvili, Nicholas Polson, and Vadim Sokolov.
\newblock Generative {{Causal Inference}}, June 2023{\natexlab{b}}.

\bibitem[Ostrovski et~al.(2018)Ostrovski, Dabney, and Munos]{ostrovski2018autoregressive}
Georg Ostrovski, Will Dabney, and R{\'e}mi Munos.
\newblock Autoregressive {{Quantile Networks}} for {{Generative Modeling}}, June 2018.

\bibitem[Padilla et~al.(2022)Padilla, Tansey, and Chen]{padilla2022quantile}
Oscar Hernan~Madrid Padilla, Wesley Tansey, and Yanzhen Chen.
\newblock Quantile regression with {{ReLU}} networks: Estimators and minimax rates.
\newblock \emph{The Journal of Machine Learning Research}, 23\penalty0 (1):\penalty0 247:11251--247:11292, January 2022.

\bibitem[Parzen(2004)]{parzen2004quantile}
Emanuel Parzen.
\newblock Quantile {{Probability}} and {{Statistical Data Modeling}}.
\newblock \emph{Statistical Science}, 19\penalty0 (4):\penalty0 652--662, 2004.

\bibitem[Pearl(2009)]{pearl2009causality}
Judea Pearl.
\newblock \emph{Causality}.
\newblock Cambridge university press, 2009.

\bibitem[Polson et~al.(2021)Polson, Sokolov, and Xu]{polson2021deep}
Nicholas Polson, Vadim Sokolov, and Jianeng Xu.
\newblock Deep learning partial least squares.
\newblock \emph{arXiv preprint arXiv:2106.14085}, 2021.

\bibitem[Polson and Ro{\v c}kov{\'a}(2018)]{polson2018posterior}
Nicholas~G Polson and Veronika Ro{\v c}kov{\'a}.
\newblock Posterior {{Concentration}} for {{Sparse Deep Learning}}.
\newblock In \emph{Advances in {{Neural Information Processing Systems}}}, volume~31. Curran Associates, Inc., 2018.

\bibitem[Polson and Sokolov(2023)]{polson2023generative}
Nicholas~G. Polson and Vadim Sokolov.
\newblock Generative {{AI}} for {{Bayesian Computation}}, June 2023.

\bibitem[Polson et~al.(2024)Polson, Ruggeri, and Sokolov]{polson2024generative}
Nick Polson, Fabrizio Ruggeri, and Vadim Sokolov.
\newblock Generative {{Bayesian Computation}} for {{Maximum Expected Utility}}.
\newblock \emph{Entropy}, 26\penalty0 (12):\penalty0 1076, December 2024.

\bibitem[Puelz et~al.(2022)Puelz, Basse, Feller, and Toulis]{puelz2022graphtheoretic}
David Puelz, Guillaume Basse, Avi Feller, and Panos Toulis.
\newblock A {{Graph-Theoretic Approach}} to {{Randomization Tests}} of {{Causal Effects}} under {{General Interference}}.
\newblock \emph{Journal of the Royal Statistical Society Series B: Statistical Methodology}, 84\penalty0 (1):\penalty0 174--204, February 2022.

\bibitem[Rosenbaum and Rubin(1983)]{rosenbaum1983central}
Paul~R Rosenbaum and Donald~B Rubin.
\newblock The central role of the propensity score in observational studies for causal effects.
\newblock \emph{Biometrika}, 70\penalty0 (1):\penalty0 41--55, 1983.

\bibitem[Rubin(1974)]{rubin1974estimating}
Donald~B. Rubin.
\newblock Estimating causal effects of treatments in randomized and nonrandomized studies.
\newblock \emph{Journal of Educational Psychology}, 66\penalty0 (5):\penalty0 688--701, October 1974.

\bibitem[Rubin and Waterman(2006)]{rubin2006estimating}
Donald~B. Rubin and Richard~P. Waterman.
\newblock Estimating the {{Causal Effects}} of {{Marketing Interventions Using Propensity Score Methodology}}.
\newblock \emph{Statistical Science}, 21\penalty0 (2):\penalty0 206--222, 2006.

\bibitem[{Schmidt-Hieber}(2020)]{schmidt-hieber2020nonparametric}
Johannes {Schmidt-Hieber}.
\newblock Nonparametric regression using deep neural networks with {{ReLU}} activation function.
\newblock \emph{The Annals of Statistics}, 48\penalty0 (4):\penalty0 1875--1897, August 2020.

\bibitem[{Schmidt-Hieber} and Zamolodtchikov(2024)]{schmidt-hieber2024generative}
Johannes {Schmidt-Hieber} and Petr Zamolodtchikov.
\newblock Generative {{Modelling}} via {{Quantile Regression}}, September 2024.

\bibitem[Shen et~al.(2021)Shen, Jiao, Lin, Horowitz, and Huang]{shen2021deep}
Guohao Shen, Yuling Jiao, Yuanyuan Lin, Joel~L. Horowitz, and Jian Huang.
\newblock Deep {{Quantile Regression}}: {{Mitigating}} the {{Curse}} of {{Dimensionality Through Composition}}, July 2021.

\bibitem[{Sohl-Dickstein} et~al.(2015){Sohl-Dickstein}, Weiss, Maheswaranathan, and Ganguli]{sohl-dickstein2015deep}
Jascha {Sohl-Dickstein}, Eric~A. Weiss, Niru Maheswaranathan, and Surya Ganguli.
\newblock Deep {{Unsupervised Learning}} using {{Nonequilibrium Thermodynamics}}, November 2015.

\bibitem[{Splawa-Neyman} et~al.(1990){Splawa-Neyman}, Dabrowska, and Speed]{splawa-neyman1990application}
Jerzy {Splawa-Neyman}, D.~M. Dabrowska, and T.~P. Speed.
\newblock On the {{Application}} of {{Probability Theory}} to {{Agricultural Experiments}}. {{Essay}} on {{Principles}}. {{Section}} 9.
\newblock \emph{Statistical Science}, 5\penalty0 (4), November 1990.

\bibitem[Stephens et~al.(2023)Stephens, Nobre, Moodie, and Schmidt]{stephens2023causal}
David~A. Stephens, Widemberg~S. Nobre, Erica E.~M. Moodie, and Alexandra~M. Schmidt.
\newblock Causal {{Inference Under Mis-Specification}}: {{Adjustment Based}} on the {{Propensity Score}}.
\newblock \emph{Bayesian Analysis}, -1\penalty0 (-1):\penalty0 1--46, January 2023.

\bibitem[Taddy et~al.(2016)Taddy, Gardner, Chen, and Draper]{taddy2016nonparametric}
Matt Taddy, Matt Gardner, Liyun Chen, and David Draper.
\newblock A {{Nonparametric Bayesian Analysis}} of {{Heterogenous Treatment Effects}} in {{Digital Experimentation}}.
\newblock \emph{Journal of Business \& Economic Statistics}, 34\penalty0 (4):\penalty0 661--672, October 2016.

\bibitem[Telgarsky(2015)]{telgarsky2015representation}
Matus Telgarsky.
\newblock Representation {{Benefits}} of {{Deep Feedforward Networks}}, September 2015.

\bibitem[Telgarsky(2016)]{telgarsky2016benefits}
Matus Telgarsky.
\newblock Benefits of depth in neural networks.
\newblock In \emph{Conference on {{Learning Theory}}}, pages 1517--1539. PMLR, June 2016.

\bibitem[Tian and Pearl(2013)]{tian2013probabilities}
Jin Tian and Judea Pearl.
\newblock Probabilities of {{Causation}}: {{Bounds}} and {{Identification}}, January 2013.

\bibitem[Vasilescu(2022)]{vasilescu2022causal}
M.~Alex~O. Vasilescu.
\newblock Causal {{Deep Learning}}: {{Causal Capsules}} and {{Tensor Transformers}}, December 2022.

\bibitem[Wager and Athey(2018)]{wager2018estimation}
Stefan Wager and Susan Athey.
\newblock Estimation and {{Inference}} of {{Heterogeneous Treatment Effects}} using {{Random Forests}}.
\newblock \emph{Journal of the American Statistical Association}, 113\penalty0 (523):\penalty0 1228--1242, July 2018.

\bibitem[Wang and Ro{\v c}kov{\'a}(2023)]{wang2023adversarial}
Yuexi Wang and Veronika Ro{\v c}kov{\'a}.
\newblock Adversarial {{Bayesian Simulation}}, July 2023.

\bibitem[Wang et~al.(2022)Wang, Polson, and Sokolov]{wang2022data}
Yuexi Wang, Nicholas Polson, and Vadim~O Sokolov.
\newblock Data augmentation for bayesian deep learning.
\newblock \emph{Bayesian Analysis}, 1\penalty0 (1):\penalty0 1--29, 2022.

\bibitem[White(1989)]{white1989asymptotic}
Halbert White.
\newblock Some {{Asymptotic Results}} for {{Learning}} in {{Single Hidden-Layer Feedforward Network Models}}.
\newblock \emph{Journal of the American Statistical Association}, 84\penalty0 (408):\penalty0 1003--1013, December 1989.

\bibitem[White(1992)]{white1992nonparametric}
Halbert White.
\newblock Nonparametric {{Estimation}} of {{Conditional Quantiles Using Neural Networks}}.
\newblock In Connie Page and Raoul LePage, editors, \emph{Computing {{Science}} and {{Statistics}}}, pages 190--199, New York, NY, 1992. Springer.
\newblock ISBN 978-1-4612-2856-1.

\bibitem[Xu et~al.(2018)Xu, Daniels, and Winterstein]{xu2018bayesian}
Dandan Xu, Michael~J. Daniels, and Almut~G. Winterstein.
\newblock A {{Bayesian}} nonparametric approach to causal inference on quantiles: {{A Bayesian Nonparametric Approach}} to {{Causal Inference}} on {{Quantiles}}.
\newblock \emph{Biometrics}, 74\penalty0 (3):\penalty0 986--996, September 2018.

\bibitem[Zaidi and Mukherjee(2018)]{zaidi2018gaussian}
Abbas Zaidi and Sayan Mukherjee.
\newblock Gaussian {{Process Mixtures}} for {{Estimating Heterogeneous Treatment Effects}}, December 2018.

\end{thebibliography}
\end{document}